\documentclass[aps,prb,twocolumn,notitlepage,showpacs,superscriptaddress,10pt]{revtex4-1}%
\usepackage{graphicx}
\usepackage{amsmath}
\usepackage{bm}
\usepackage{amssymb}
\usepackage{color,soul}
\usepackage{amsfonts}%
\usepackage[caption=false]{subfig}
\usepackage{comment}
\usepackage{color,soul}
\usepackage[dvipsnames]{xcolor}
\usepackage[capitalize]{cleveref}
\usepackage{float}
\usepackage{tabularx}
\usepackage{array}   
\newcolumntype{L}{>{$}l<{$}} 
\newcolumntype{R}{>{$}r<{$}} 
\newcolumntype{C}{>{$}c<{$}} 
\usepackage{comment}
\usepackage[capitalize]{cleveref}

\usepackage[utf8]{inputenc}

\begin{document}

\title{IrF$_4$: From Tetrahedral Compass Model to Topological Semimetal}

\author{Chao Shang}%
\affiliation{Department of Physics and Center for Functional Materials, Wake Forest University, NC 27109, USA}

\author{Owen Ganter}%
\affiliation{Department of Physics and Center for Functional Materials, Wake Forest University, NC 27109, USA}

\author{Niclas Heinsdorf}
\affiliation{Max-Planck-Institut für Festkörperforschung, Heisenbergstrasse 1, D-70569 Stuttgart, Germany}
\affiliation{Department of Physics and Astronomy \& Stewart Blusson Quantum Matter Institute,
University of British Columbia, Vancouver BC, Canada V6T 1Z4}

\author{Stephen M. Winter}%
\affiliation{Department of Physics and Center for Functional Materials, Wake Forest University, NC 27109, USA}

\date{\today}% It is always \today, today,

\begin{abstract}
The intersection of topology, symmetry, and magnetism yields a rich structure of possible phases. In this work, we study theoretically the consequences of magnetism on IrF$_4$, which was recently identified as a possible candidate topological nodal chain semimetal in the absence of magnetic order. We show that the spin-orbital nature of the Ir moments gives rise to strongly anisotropic magnetic couplings resembling a tetrahedral compass model on a diamond lattice. The predicted magnetic ground state preserves key symmetries protecting the nodal lines, such that they persist into the ordered phase at the mean-field level. The consequences for other symmetry reductions are also discussed. 
\end{abstract}

\maketitle

\section{Introduction}

In recent years, the study of materials with strong spin-orbit coupling (SOC) has risen to prominence particularly in conjunction with recent developments on topological phases of weakly interacting electrons\cite{bansil2016colloquium}, such as topological insulators\cite{hasan2010colloquium,hasan2011three,tokura2019magnetic} and semimetals \cite{xu2015discovery,huang2015weyl,weng2015weyl,lv2015experimental,yan2017topological,armitage2018weyl}. Particularly intriguing cases occur when crystalline symmetries enrich or enforce aspects of the band topology. For example, Ref.~\onlinecite{bzduvsek2016nodal} recently showed that specific space groups with mutually orthogonal glide planes can support topologically protected nodal lines, which are linked together to form chains at high symmetry points in the Brillouin zone. 
As a proof of concept, the authors considered an idealized nearest neighbor hopping model for IrF$_4$. It was shown that this model indeed exhibits nodal chain Fermi surfaces, along with associated ``drumhead'' surface states \cite{weng2015topological}. Experimentally, little has been reported on IrF$_4$, although early studies\cite{rao1976tetrafluorides} reveal a sizeable magnetic susceptibility consistent with antiferromagnetically coupled Ir moments. The possible effects of magnetic order on the fate of the nodal chain semimetal state have yet to be explored. From previous works on magnetic interactions between heavy $d^5$ metals\cite{chen2008spin,jackeli2009mott,kim2012magnetic,winter2016challenges,bertinshaw2019square}, it can be expected that the effective magnetic couplings between Ir moments are strongly anisotropic, giving rise to potentially interesting magnetic properties.

In this context, it may be noted that various antiferromagnetic topological semimetals have been predicted and/or discovered, such as the frustrated all-in-all-out pyrochlore iridates\cite{wan2011topological,yang2011quantum,witczak2012topological,witczak2013pyrochlore,yamaji2014metallic} A$_2$Ir$_2$O$_7$, and stacked Kagome antiferromagnets\cite{kubler2014non,nakatsuji2015large,nayak2016large,kiyohara2016giant,yang2017topological,park2018magnetic} (e.g.~Mn$_3$Sn and Mn$_3$Ge). Typically, the Fermi surfaces consist of isolated Weyl or Dirac points when including both time-reversal symmetry breaking and SOC \cite{vsmejkal2017route,zou2019study}, since nodal lines are not protected in the absence of additional symmetries\cite{burkov2011topological,fang2015topological,bzduvsek2017symmetry,armitage2018weyl}. In particular, in the absence of time-reversal symmetry (TRS), nodal lines can only be guaranteed to be pinned to the Fermi Level in the presence of a chiral (sublattice) symmetry $\hat{C}$, with $\hat{C}\mathcal{H}\hat{C}^{-1} = - \mathcal{H}$. Such a symmetry is approximately realised only for specific magnetic orders on bipartite lattices, under the condition of dominant intersublattice hopping. These conditions are not satisfied for the non-bipartite pyrochlore and kagome lattices relevant to the above mentioned antiferromagnetic Weyl and Dirac semimetals. They are also violated for conventional colinear two-sublattice N\'eel antiferromagnetic order on bipartite lattices, where the symmetry breaking antiferromagnetic term can be shown to commute with $\hat{C}$. For these reasons, {\it antiferromagnetic} nodal line semimetals (AF-NLSMs) are notably rare. 

In general, the search for topological phases of matter and surface modes (e.g. as they appear in paramagnetic IrF$_4$) relies on classification schemes that are based on crystalline and non-spatial symmetries of a given material\cite{fu2011topological, schnyder2008classification, bradlyn2017topological, elcoro2021magnetic,po2017symmetry,tang2019comprehensive,khalaf2018symmetry}. Symmetries that are only present at lower-order approximations of an effective model are called \textit{quasi-symmetries}. A surface mode is said to be protected by a quasi-symmetry if the higher-order perturbation is small compared to the gap in which these states are located\cite{guo2022quasi, hu2022hierarchy}. Typically, the discovery and prediction of surface modes protected by quasi-symmetries is difficult since the (magnetic) space groups of the lower-symmetry effective models alone give no indication to the presence of topological phases. Instead, the system inherits the surface modes and transport properties from the higher-symmetry effective model as long as the symmetry-breaking is small. Strictly, these surface modes are not topologically protected because the symmetry-breaking term gaps them out weakly such that they can be removed by an adiabatic transformation of the Hamiltonian. The crystal symmetries that are responsible for the appearance of surface modes in the paramagnetic IrF$_4$ compound become quasi-symmetries in the antiferromagnetic state with small moments, whereas the approximate chiral symmetry is a quasi-symmetry for both phases. 

In this work, we show that antiferromagnetically ordered IrF$_4$ showcases two distinct types of quasi-symmetries (crystalline and non-spatial) and uncover their effect on the vestigial surface modes that are passed down from the topological, paramagnetic parent compound. Further, we reveal a possible route to an AF-NLSM. The magnetic model for IrF$_4$ is shown to approximate a strongly anisotropic tetrahedral compass model, with stripy magnetic order rather than the conventional N\'eel state. When this magnetic pattern is considered in conjunction with the nearest neighbor hopping model of Ref.~\onlinecite{bzduvsek2016nodal}, we find an AF-NLSM state survives at the mean field level due to the retention of an approximate chiral symmetry. Thus, if IrF$_4$ realises an itinerant magnetic state, it is a good candidate for an AF-NLSM. 

The paper is organized as follows. In section \ref{sec:magnetism}, we first derive a model of the magnetic interactions between Ir moments, and determine the magnetic ground state to be stripy antiferromagnetic. In section \ref{sec:bands}, we then analyze the consequences of TRS breaking magnetic orders on the bulk bands and surface states in an itinerant picture. 

\begin{figure}
\centering
\includegraphics[width=0.9\columnwidth]{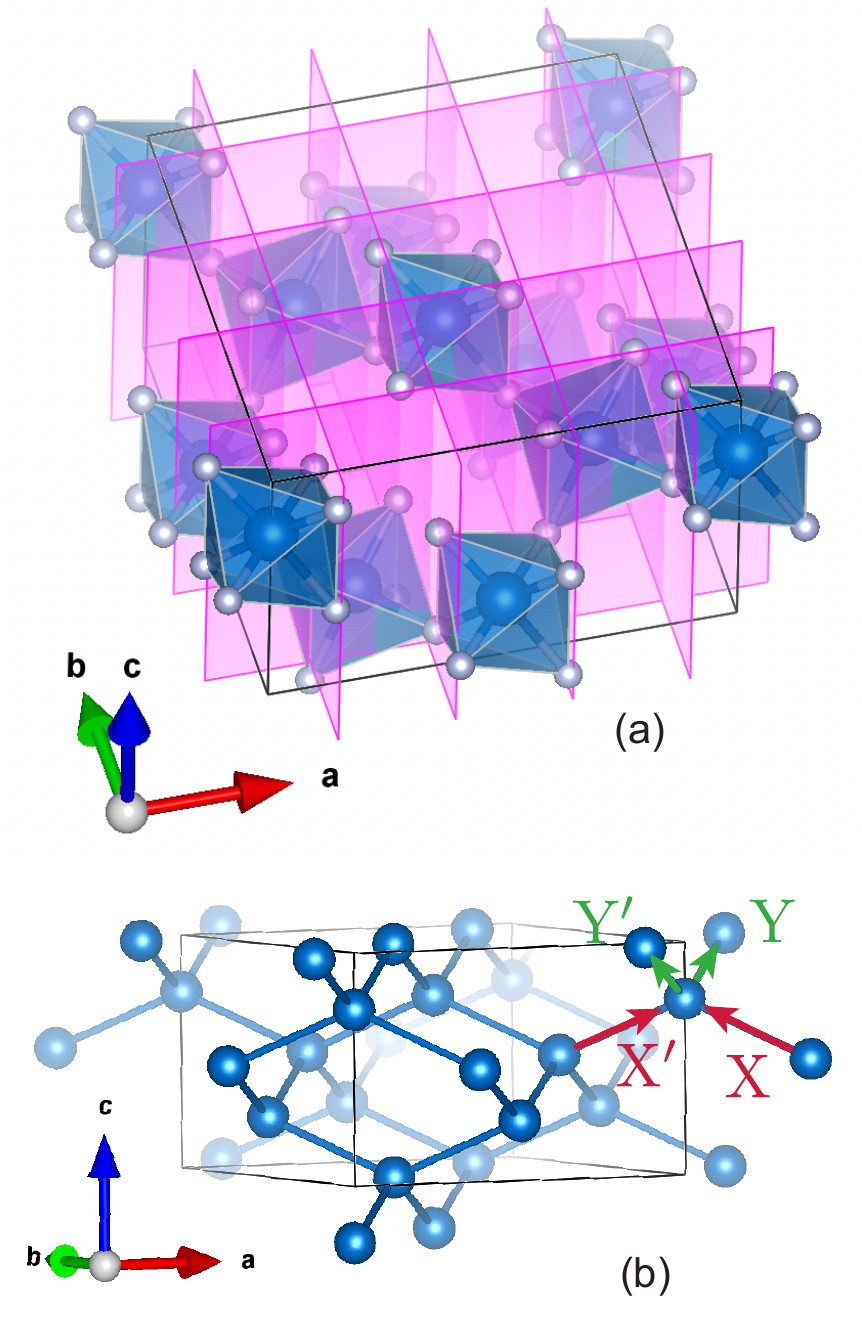}
\caption{(a) View of the $Fdd2$ unit cell of IrF$_4$, showing $d$-glide planes. (b) Nearest neighbor Ir-Ir bonds, emphasizing diamond lattice connectivity. Different types of bonds are indicated.}
\label{fig:structure}
\end{figure}

\section{Magnetism} \label{sec:magnetism}

\subsection{Methods} \label{sec:methods}
In order to estimate the magnetic interactions between Ir sites, we utilize the exact diagonalization approach outlined in Ref.~\onlinecite{winter2016challenges}. We consider pairs of Ir sites, described by the total Hamiltonian:
\begin{align}\label{eqn:totalham}
\mathcal{H} = \mathcal{H}_{\rm CFS} + \mathcal{H}_{\rm SOC} + \mathcal{H}_{U} + \mathcal{H}_{\rm hop},
\end{align}
including all five $d$-orbitals at each site. Here, $\mathcal{H}_{\rm CFS}$ is the local on-site crystal field splitting, $\mathcal{H}_{\rm SOC}$ describes the on-site spin-orbit coupling, and $\mathcal{H}_{\rm hop}$ gives the hopping between sites. 
The Coulomb interactions are most generally written:
\begin{align}
\mathcal{H}_U = \sum_i\sum_{\alpha,\beta,\delta,\gamma}\sum_{\sigma,\sigma^\prime}U_{\alpha\beta\gamma\delta} \ c_{i,\alpha,\sigma}^\dagger c_{i,\beta,\sigma^\prime}^\dagger c_{i,\gamma,\sigma^\prime} c_{i,\delta,\sigma}
\end{align}
where $i$ labels the site, $\alpha,\beta,\gamma,\delta$ label different $d$-orbitals, and $\sigma,\sigma^\prime$ indicate different spin indices. We assume spherically symmetric interactions\cite{sugano2012multiplets}, for which the coefficients $U_{\alpha\beta\gamma\delta}$ are all related to the three Slater parameters $F_0, F_2, F_4$. Throughout, we assume approximate ratio\cite{pavarini2014dmft} $F_4/F_2 = 5/8$, and parameterize the interactions via: $U_{t2g} = F_0 + \frac{4}{49} \left( F_2 + F_4 \right) = 1.7$ eV and $J_{t2g} = \frac{3}{49} F_2 + \frac{20}{441} F_4 = 0.3$ eV, which are compatible with estimates from constrained RPA\cite{yamaji2014first} as well as previous calculations on iridates\cite{winter2016challenges}. 

In order to parameterize the single-particle part of the Hamiltonian $\mathcal{H}_{\rm CFS} + \mathcal{H}_{\rm SOC} + \mathcal{H}_{\rm hop}$ (on-site crystal field, spin-orbit coupling, and intersite hopping, respectively), we performed fully relativistic DFT calculations with FPLO\cite{opahle1999full} at the GGA (PBE\cite{perdew1996generalized}) level with a 12$\times$12$\times$12 $k$-mesh based on the structure reported in Ref.~\onlinecite{jain2013commentary,osti_1189775}. The resulting Kohn-Sham eigenvectors were then projected onto Ir $d$-orbital Wannier functions \cite{koepernik2021symmetry}. Computed hoppings are given in Appendix \ref{sec:appendix}. The crystal field terms are discussed in section \ref{sec:crystalfield}.

To estimate the magnetic couplings, the combined Hamiltonian (Eq.~\ref{eqn:totalham}) was then diagonalized for two Ir sites, and the low-energy eigenstates were projected onto pure $j_{1/2}$ states to derive the low-energy magnetic Hamiltonian. Results are presented in Sec.~\ref{sec:interactions}. As discussed in Ref.~\onlinecite{winter2016challenges}, this latter step is required to establish an appropriate definition of the low-energy spin states;  the resulting magnetic couplings are not particularly sensitive to choice of projection basis, as long as the overlap between the projection basis and exact low-energy space is finite. The computed couplings are guaranteed to respect all symmetries, and represent a non-perturbative estimate that remains valid even away from the large $U/t$ limit. For this reason, we expect the local magnetic Hamiltonian to accurately describe possible magnetic orders even if the Ir electrons are relatively itinerant. 

\subsection{Crystal Field Distortions} \label{sec:crystalfield}

\begin{figure}
\centering
\includegraphics[width=1.0\columnwidth]{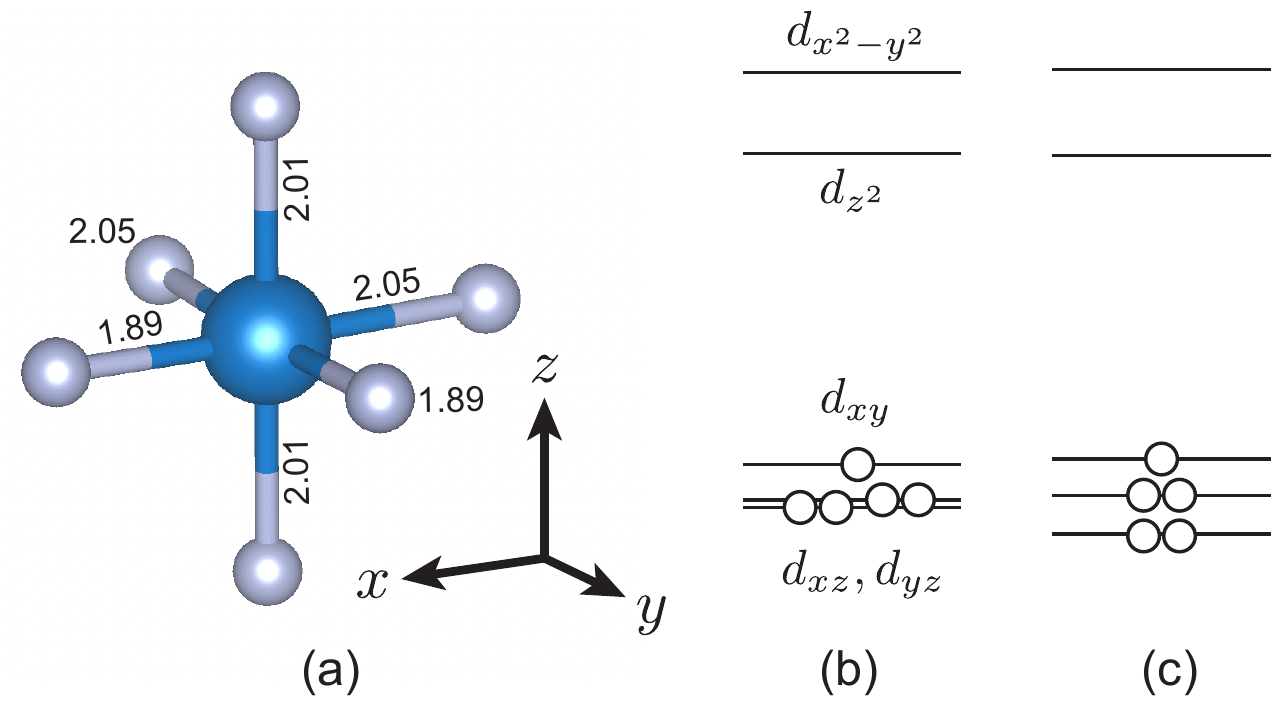}
\caption{(a) Distorted IrF$_6$ octahedra with bond lengths indicated and definition of local $(x,y,z)$ coordinates. The local $C_2$ axis is along the $\hat{x}+\hat{y}$ direction. (b) Energetic splitting of the $d$-orbitals without SOC. (c) Energy levels with SOC. Occupied states for $d^5$ filling of Ir(IV) are indicated.}
\label{fig:crystalfield}
\end{figure}

\begin{table}
\centering\def\arraystretch{1.1}
    \begin{ruledtabular}
    \caption{Computed crystal field matrix elements with orbitals defined according to the coordinates in Fig.~\ref{fig:crystalfield}(a). \label{tab:crystalfield}}
    \begin{tabular}{R|RRRRR}
    & d_{yz} & d_{xz} & d_{xy}  & d_{z^2} & d_{x^2\text{-}y^2}
   \\
    \hline
   d_{yz}  		&-0.302&		 -0.052&		 -0.030&		-0.319&		 -0.248  	 \\
d_{xz} 			&-0.052&		 -0.302&		0.030&		0.319&		 -0.248   	 \\
d_{xy}                  &-0.030&		0.030&		 -0.074&		0.048&	0		\\
d_{z^2}      &-0.319&		-0.319&		0.048&		2.180&		0	\\
d_{x^2\text{-}y^2}                 &-0.248&		 -0.248&		 0&		0&		 2.832	\\

    \end{tabular}
    \end{ruledtabular}
    
    \label{tab:hoppings_rel}
\end{table}

In IrF$_4$, each Ir site occupies a distorted octahedral coordination environment, with local $C_2$ symmetry axis along the local cubic $x+y$ direction defined in Fig.~\ref{fig:crystalfield}(a). The original reported structure based on powder x-ray analysis\cite{bartlett1974} suggested significant distortion of the local IrF$_6$ octahedra, with Ir$-$F bond lengths ranging from 1.89~\AA \ to 2.08~\AA. Optimization of the structure using density functional theory at the GGA level\cite{jain2013commentary,osti_1189775} yield somewhat reduced distortions, but the Ir$-$F bonds still show significant anisotropy, ranging between 1.89~\AA \ and 2.05~\AA, as shown in Fig~\ref{fig:crystalfield}(a). All subsequent calculations have been performed on the optimized structure of Ref.~\onlinecite{osti_1189775}. The local distortion is reflected in the computed crystal field parameters, summarized in Table \ref{tab:crystalfield}. 

In the absence of SOC, the $t_{2g}$ orbitals are split into a nearly degenerate $d_{xz},d_{yz}$ pair and higher lying $d_{xy}$ orbital (Fig~\ref{fig:crystalfield}(b)). The splitting is of the order of $0.23$ eV, which is competitive with the spin-orbit coupling, described approximately by $\lambda \mathbf{L} \cdot \mathbf{S}$, with $\lambda = 0.4$ eV. As a result, when SOC is included, the unpaired electron in the $t_{2g}$ manifold does not occupy a pure $j_{\rm eff} = \frac{1}{2}$ state. Instead, the low-energy doublet is approximately described by:
\begin{align} \label{eqn:doublet1}
    |+\rangle =& \  \alpha |xy,\uparrow\rangle + \beta(i|xz,\downarrow\rangle + |yz,\downarrow\rangle)\\ \label{eqn:doublet2}
    |-\rangle =& \  -\alpha |xy,\downarrow\rangle +\beta(-i|xz,\uparrow\rangle + |yz,\uparrow\rangle)
\end{align}
with $\alpha \approx 0.77$ and $\beta \approx 0.45$ reflecting larger weight in the $d_{xy}$ orbitals. Nonetheless, the low-energy doublet retains large overlap with a pure $j_{1/2}$ state (for which $\alpha = \beta = 1/\sqrt{3}$), with $|\langle j_{1/2}, \uparrow| +\rangle| \approx 0.96$. Thus, in the following, we expect no errors in the analysis of the magnetic couplings from projecting onto pure $j_{1/2}$ states.

\subsection{Magnetic Hamiltonian}
\label{sec:interactions}

\begin{figure}
\centering
\includegraphics[width=\columnwidth]{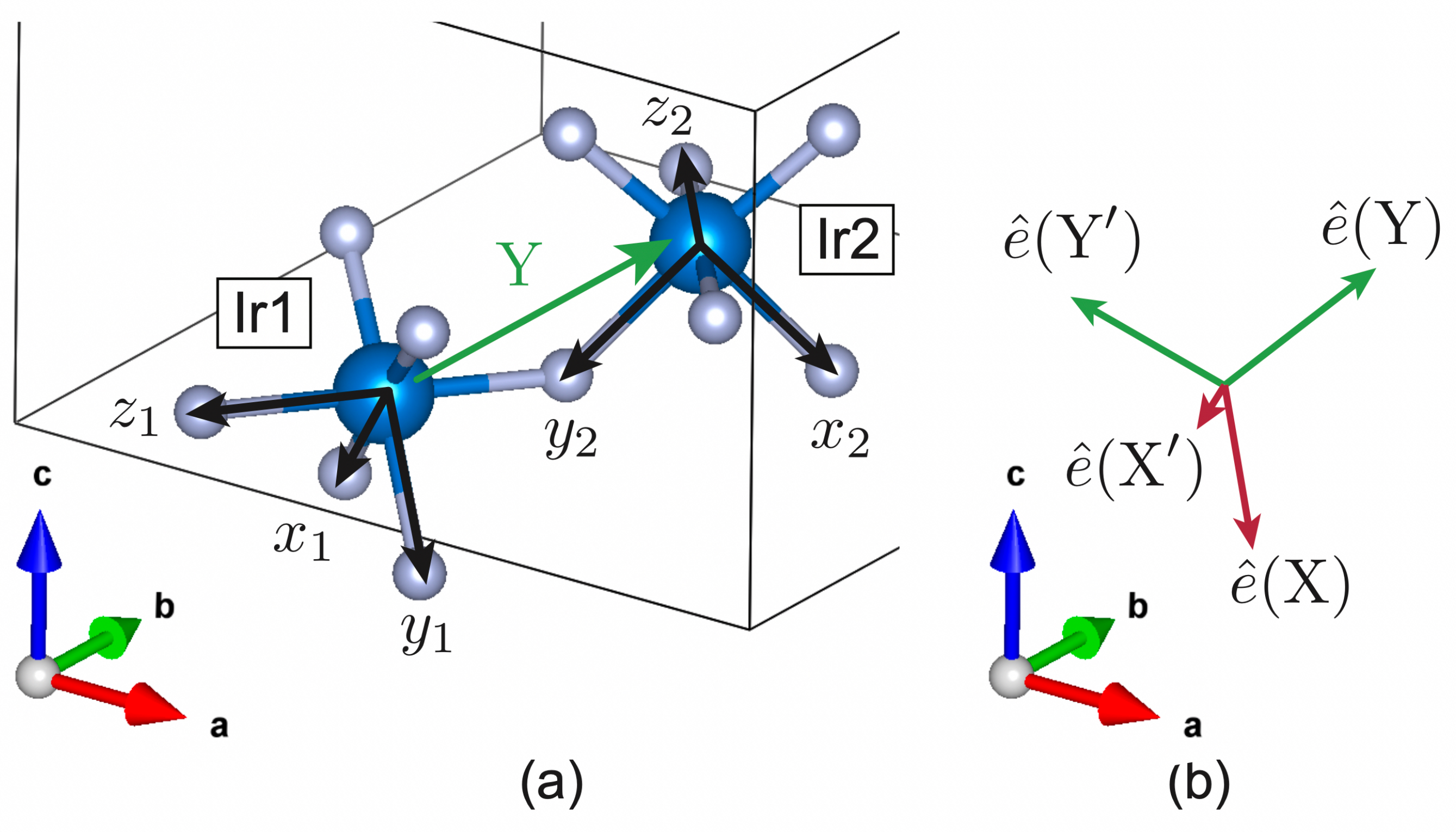}
\caption{(a) Geometry of the Y-bond with local coordinates indicated for each Ir site, in accordance with Fig.~\ref{fig:crystalfield}(a). (b) Tetrahedral vectors $\hat{e}$ defining the idealized interactions along each nearest neighbor bond.}
\label{fig:bond}
\end{figure}

In the IrF$_4$ structure, there are four distinct types of nearest neighbor bonds, labelled X, X$^\prime$, Y, and Y$^\prime$ in Fig.~\ref{fig:structure}(b). We first focus on the Y-bond depicted in Fig.~\ref{fig:bond}(a). The computed magnetic interactions are described by:
\begin{align}
    \mathcal{H}_{\text{Y-bond}} = \mathbf{S}_1 \cdot \mathbb{J}_{12}^{\rm Y} \cdot \mathbf{S}_2
\end{align}
where the spins $\mathbf{S}_1$ and $\mathbf{S}_2$ correspond to Ir1 and Ir2 in Fig.~\ref{fig:bond}(a). The interaction tensor is:
\begin{align}
    \mathbb{J}_{12}^{\rm Y} = \left( \begin{array}{ccc}
    -6.9 & 14.9 & 11.4 \\
    9.7 & -5.7 & 17.4 \\
    14.2 & 9.0 & -6.5
    \end{array}
    \right) \text{ meV}
\end{align}
in terms of the global crystallographic $(a,b,c)$ coordinates. In order to interpret these couplings, it is convenient to write $\mathbb{J}$ in terms of the bond-Hamiltonian $\mathcal{H}_{12} = J_{12} \ \mathbf{S}_1 \cdot \mathbf{S}_2 + \mathbf{D}_{12} \cdot (\mathbf{S}_1 \times \mathbf{S}_2) + \mathbf{S}_1 \cdot \Gamma \cdot \mathbf{S}_2$, where:
\begin{align}
    J_{12} = (1/3)\text{Tr} [\mathbb{J}_{12}^{\rm Y} ] = -6.4\text{ meV}
\end{align}
is the isotropic Heisenberg coupling, 
\begin{align}
    \mathbf{D}_{12}^{\rm Y} = (-4.2, -1.4, -2.6)\text{ meV}
\end{align}
parameterizes the antisymmetric Dzyalloshinkii-Moriya (DM) interactions, and:
\begin{align}
    \Gamma_{12}^{\rm Y} = & \ \frac{1}{2} \left[ \mathbb{J}_{12}^{\rm Y} + (\mathbb{J}_{12}^{\rm Y} )^T\right] -  J_{12}\ \mathbb{I}_{3\times 3} \\
    = & \ \left( \begin{array}{ccc}
    -0.5 & 12.3 & 12.7 \\
    12.3 & 0.6 & 13.2 \\
    12.7 & 13.2 & -0.1
    \end{array}
    \right) \text{ meV}
\end{align}
is the traceless symmetric pseudo-dipolar tensor. Interactions for the remaining bonds may be found via symmetry. The Y$^\prime$-bonds are related to the Y-bonds via twofold rotation along the $c$-axis, such that:
\begin{align}
    \mathbf{D}_{12}^{\rm Y^\prime} = & \  (+4.2,+1.4,-2.6) \text{ meV}\\
    \Gamma_{12}^{\rm Y^\prime } = & \ \left( \begin{array}{ccc}
    -0.5 & 12.3 & -12.7 \\
    12.3 & 0.6 & -13.2 \\
   - 12.7 & -13.2 & -0.1
    \end{array}
    \right) \text{ meV}
\end{align}
Similarly, the X-bond is related to the Y-bond by $d$-glide perpendicular to the $a$-axis. This gives:
\begin{align}
    \mathbf{D}_{12}^{\rm X} = & \  (-4.2,+1.4,+2.6) \text{ meV}\\
    \Gamma_{12}^{\rm X } = & \ \left( \begin{array}{ccc}
    -0.5 & -12.3 & -12.7 \\
    -12.3 & 0.6 & 13.2 \\
   - 12.7 & 13.2 & -0.1
    \end{array}
    \right) \text{ meV}
\end{align}
and subsequently:
\begin{align}
    \mathbf{D}_{12}^{\rm X^\prime} = & \  (+4.2,-1.4,+2.6) \text{ meV}\\
    \Gamma_{12}^{\rm X^\prime} = & \ \left( \begin{array}{ccc}
    -0.5 & -12.3 & 12.7 \\
    -12.3 & 0.6 & -13.2 \\
   12.7 & -13.2 & -0.1
    \end{array}
    \right) \text{ meV}
\end{align}
We therefore find that the magnetic interactions are strongly anisotropic and bond-dependent. If we approximate the small diagonal elements of $\Gamma_{12}$ to be zero, and the off-diagonal elements of $\Gamma_{12}$ to be equal in magnitude, then the symmetric interactions (Heisenberg and pseudo-dipolar) can be approximately summarized in terms of bond-dependent Ising couplings:
\begin{align} \label{eqn:compass}
    \mathcal{H}_{\rm sym} = \sum_{ij} J \ \mathbf{S}_i \cdot \mathbf{S}_j + \Gamma \ (\mathbf{S}_i \cdot \hat{e}_{ij}) (\mathbf{S}_j \cdot \hat{e}_{ij}) 
\end{align}
where $J = -19.1$ meV, $\Gamma = +38.3$ meV $\approx -2J$, and the unit vectors are:
\begin{align}
    \hat{e}_{ij}(\text{Y}) =& \ \frac{1}{\sqrt{3}}(1,1,1)\\
    \hat{e}_{ij}(\text{Y}^\prime) =& \ \frac{1}{\sqrt{3}}(-1,-1,1)\\
    \hat{e}_{ij}(\text{X}) =& \ \frac{1}{\sqrt{3}}(1,-1,-1)\\
    \hat{e}_{ij}(\text{X}^\prime) =& \ \frac{1}{\sqrt{3}}(-1,1,-1)
\end{align}
These vectors may be recognized as the tetrahedral vectors depicted in Fig.~\ref{fig:bond}(b). Therefore, IrF$_4$ approximately realises a {\it tetrahedral} quantum compass model on a diamond lattice. A similar model was recently proposed\cite{nikolaev2018realization} for CuAl$_2$O$_4$, although it was subsequently found that Jahn-Teller distortions ruin the $j_{1/2}$ ground state\cite{huang2022resonant}. In the case of IrF$_4$, the perfect correspondence with the compass model is spoiled by the DM interactions, which are significantly smaller in magnitude than $J$ and $\Gamma$, with $|\mathbf{D}| \approx 5$ meV. In the following, we show that the DM interaction may be ignored in first approximation, and focus on the compass model defined by Eq. \ref{eqn:compass}. 

\begin{figure}[b]
\centering
\includegraphics[width=\columnwidth]{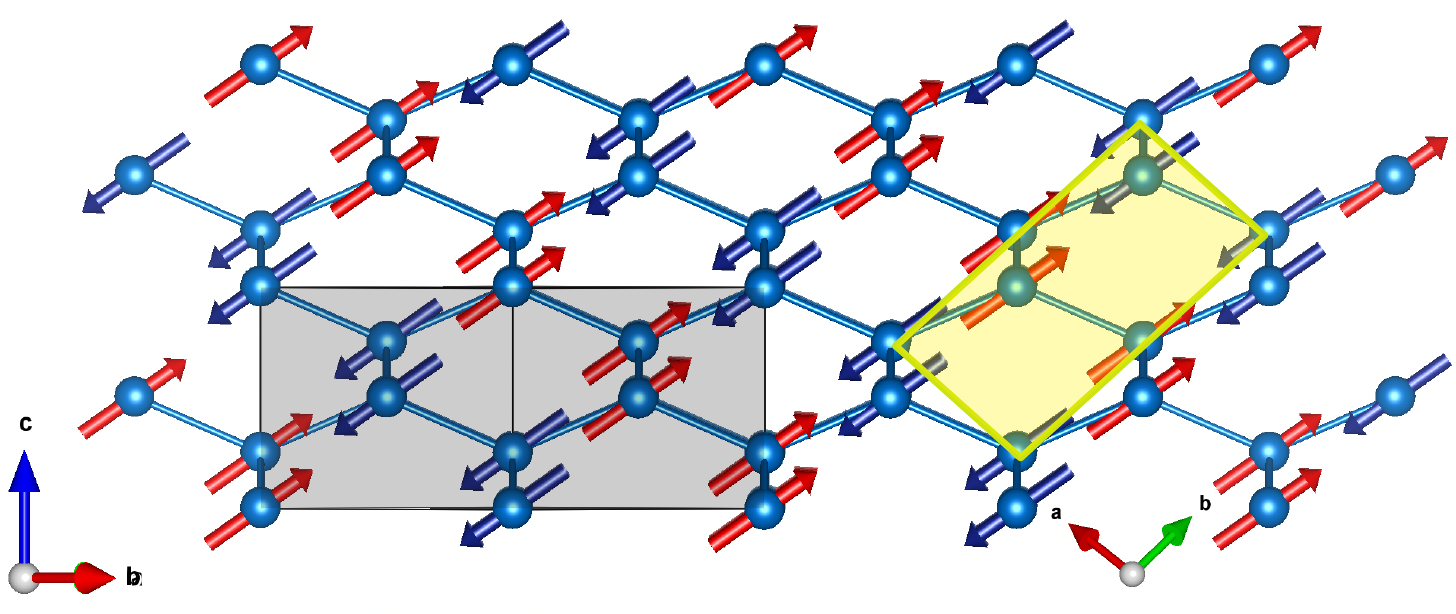}
\caption{Layered zigzag ground state magnetic structure of $\mathcal{H}_{\rm sym}$ (Eq.~\ref{eqn:compass}) corresponding to spins oriented along the $\pm \hat{e}$(Y) directions. The black unit cell is the non-magnetic $Fdd2$ unit cell. The yellow unit cell is the magnetic $P_{2s}1$ unit cell.
}
\label{fig:magstructure}
\end{figure}

\subsection{Magnetic Order}
In order to identify possible magnetic ordering patterns for IrF$_4$, we performed both Luttinger-Tisza\cite{kaplan2007spin} analysis and classical Monte Carlo simulated annealing on with $\mathcal{H}_{\rm sym}$ (Eq. \ref{eqn:compass}) as implemented in SpinW\cite{toth2015linear} for system sizes up to 10$\times$10$\times$10 orthorhombic unit cells (8000 spins). 
There are four degenerate ordering wavectors, one of which is depicted in Fig.~\ref{fig:magstructure}.
The pictured order is composed of colinear spins oriented along the $\pm \hat{e}$(Y) direction. Sites linked by the Y-bonds belong to different magnetic sublattices, as the ordered moment direction corresponds to the antiferromagnetic Ising axis of the Y-bond interactions. The spins linked by the remaining three bond types are ferromagnetically aligned, leading to alternating ferromagnetic layers.  It can be easily seen that these magnetic orders minimize both the antiferromagnetic Ising interaction, and ferromagnetic Heisenberg coupling. The three other domains (with differing ordering wavevectors) are similarly obtained: spins are oriented along one of the $\pm\hat{e}$ axes, with spins linked by the corresponding \mbox{X-,} X$^\prime$, or Y$^\prime$ bonds belonging to different sublattices.

In order to check whether this magnetic order is stable against addition of the small estimated DM-interaction, we also performed Luttinger-Tisza analysis and simulated annealing minimization including the DM-interaction. We found that the ordering wavevector is unchanged and the lowest energy structure remains essentially colinear with little change. This result can be understood from the fact that the various orientations of the DM-vectors compete with eachother, so that there is no canted or incommensurate structure that simultaneously minimizes all DM-interactions. In addition, canting of the moments is generally disfavored by the strong Ising coupling along the antiferromagnetic bonds. As a result, the magnetic structure remains robust against such perturbations, and the DM-interactions may be neglected in the minimal model.

The obtained magnetic structure corresponds to the magnetic spacegroup $P_{2s}1$. Importantly, we find that the $d$-glide symmetry (see Fig.~\ref{fig:structure}(a)) of the paramagnetic $Fdd2$ group is broken. The only retained crystallographic symmetry is $\hat{R} \cdot \hat{g}^2$, where $\hat{R}$ is time-reversal,  $\hat{g}$ is the $d$-glide, and $\hat{g}^2$ is simple translation along the face diagonals $(\frac{1}{2} \ 0 \ \frac{1}{2})$ or $(0 \ \frac{1}{2} \ \frac{1}{2})$. The consequences of this symmetry reduction on the nodal chains is discussed in the next sections.

\section{Consequences of Magnetism on Band Structure} \label{sec:bands}

\subsection{Symmetry Analysis}

\begin{figure}
\centering
\includegraphics[width=\columnwidth]{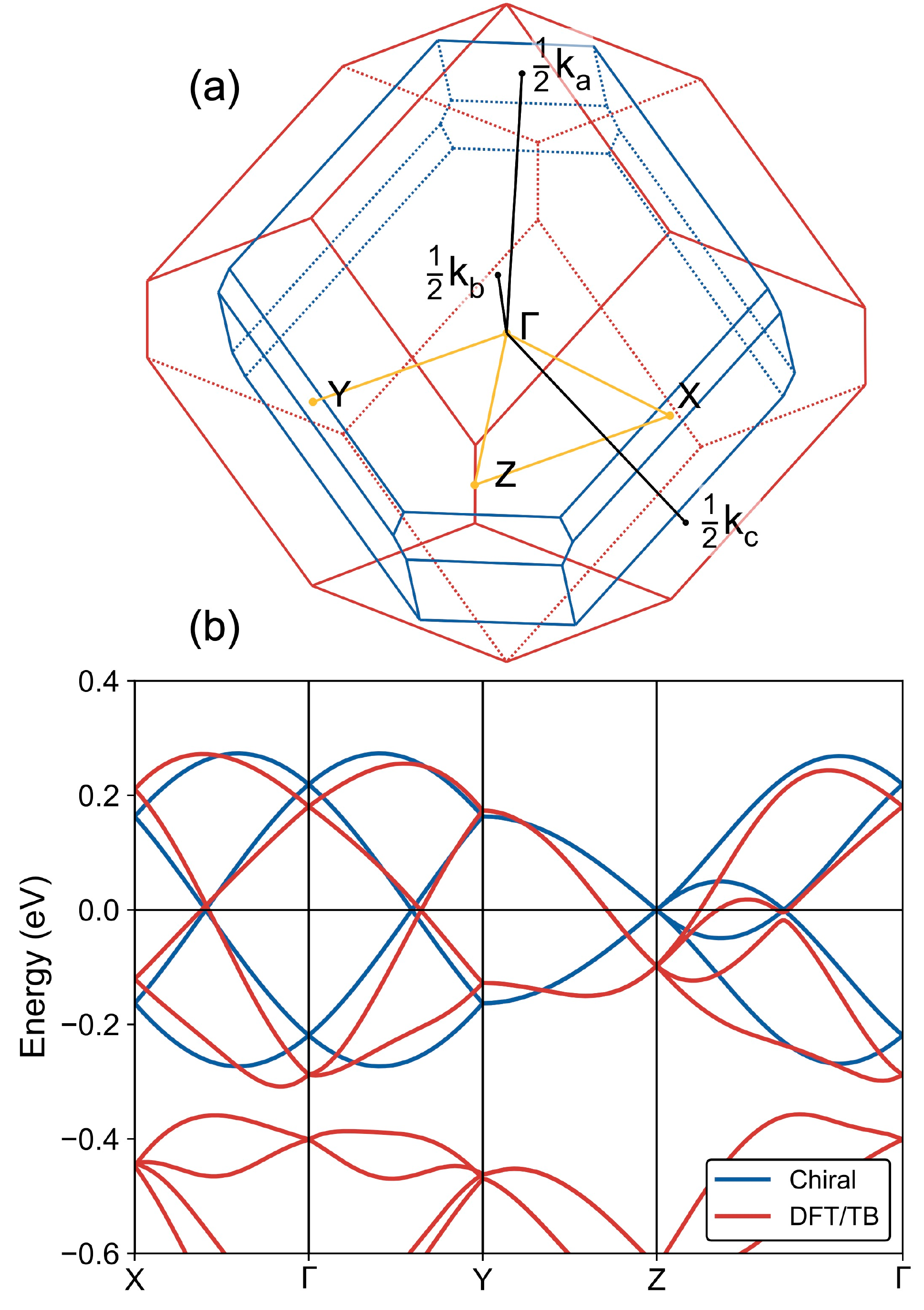}
\caption{(a) Brillouin zone of IrF$_4$ depicting a selection high-symmetry TRIM points in the non-magnetic cell (red) and magnetic cell (blue). (b) Band structure of model Eq'n (\ref{eqn:hops} compared with the ab-initio band structure including nearest neighbor hoppings.
}
\label{fig:bzero_bands}
\end{figure}

It is first insightful to review the symmetries protecting the band crossing points in the paramagnetic $Fdd2$ spacegroup, following Ref.~\onlinecite{bzduvsek2016nodal}. The primitive cell contains the two Ir sites depicted in Fig.~\ref{fig:bond}, which are related to one another by two non-symmorphic $d$-glide operations $\hat{g}_a,\hat{g}_b$, in which the glide planes are normal to the $a, b$ axes (as depicted in Fig.~\ref{fig:structure}(a)). These glides consist of a mirror operation combined with a fractional translation $\vec{d}$ by half a primitive lattice vector. In conjunction with time-reversal symmetry $\hat{R} = i\sigma_y\mathcal{K}$, it is well known\cite{young2015dirac,zhao2016nonsymmorphic,chen2016topological,leonhardt2021symmetry} that such symmetries enforce band crossings along any path in $k$-space that retains glide symmetry and connects time-reversal invariant momentum (TRIM) points $k_0$ and $k_1$ whose difference satisfies $(k_0-k_1)\cdot \vec{d} = n\pi$, where $n \in$ odd. The locations of some of such TRIM points are depicted in Fig.~\ref{fig:bzero_bands}(a). The reason for the enforced crossing is that the eigenvalues of $\hat{g}$ are given by $\pm e^{i k \cdot \vec{d}}$. At TRIM points where $k\cdot \vec{d} = n\pi$ with $n \in$ even, the eigenvectors must fall into Kramers degenerate pairs with glide eigienvalues $(+1,+1)$ or $(-1,-1)$. At TRIM points where $k\cdot \vec{d} = n\pi$ with $n \in$ odd, the Kramers pairs instead have glide eigenvalues $(+i,-i)$. As a result, a symmetry enforced crossing of bands with different glide eigenvalues is required to occur at an intermediate $k$-point.

In principle, the enforced band crossings are not required to occur at the Fermi energy. However, for IrF$_4$, if one considers only the $|+\rangle, |-\rangle$ bands with nearest neighbor hopping (which is 3.6 times larger than further neighbor hopping), the model has an additional chiral sublattice symmetry $\hat{C}$ that satisfies $\hat{C}\mathcal{H}_{\rm hop}\hat{C}^{-1} = -\mathcal{H}_{\rm hop}$. Following Ref.~\onlinecite{bzduvsek2016nodal}, $\mathcal{H}_{\rm hop}$ may be written:
\begin{align}\label{eqn:hops}
    \mathcal{H}_{\rm hop} = \sum_{\langle ij \rangle} \mathbf{c}_i^\dagger
    \left[ t_1 \mathbb{I}_{2\times 2} + i T_1 (\hat{r}_{ij}\times \hat{e}_Z) \cdot \vec{\sigma}  \right] \mathbf{c}_j
\end{align}
where $\hat{e}_Z$ is a unit vector along the $c$-axis of the paramagnetic $Fdd2$ cell. $\hat{C}$ corresponds to taking $\mathbf{c} \to -\mathbf{c}$ for one of the crystallographic sublattices (e.g. Ir2 in Fig.~\ref{fig:bond}(a)). DFT calculations in Ref.~\onlinecite{bzduvsek2016nodal} provided estimates of $t_1 =  0.0548$ and $T_1 = -0.0577$ eV; these are compatible with our results as well. The band structure of the paramagnetic model eq'n (\ref{eqn:hops}) is compared in Fig.~\ref{fig:bzero_bands}(b) to the results of including the full nearest neighbor hoppings computed via DFT (see Sec.~\ref{sec:methods}). Symmetry enforced crossings occur, for example, along the $\Gamma \to \text{X}$ and $\Gamma \to \text{Y}$ paths. 

\begin{figure*}
\centering
\includegraphics[width=\linewidth]{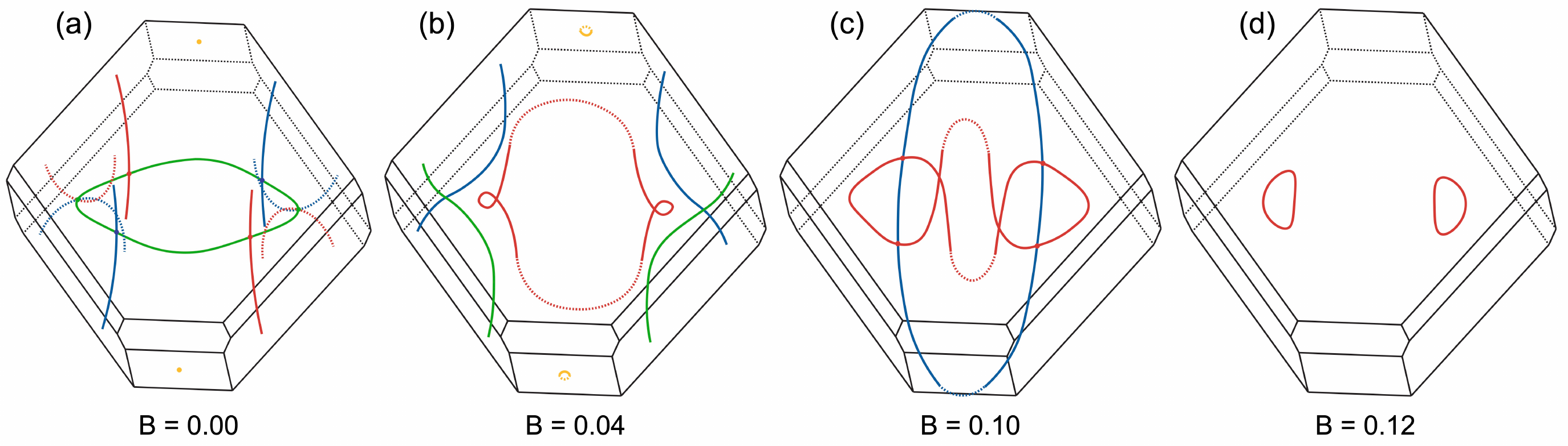}
\caption{Evolution of nodal line Fermi surface as a function of staggered zigzag mean field $B_{\rm ZZ}$ for the model of Eq'n (\ref{eqn:h1p}). In (a), dashed lines indicate nodal lines folded into the magnetic Brillouin zone. In (b) and (c), dashed lines indicate sections of the Fermi surface that have been translated to higher Brillouin zones for clarity of connectivity. 
The pictured fields correspond to staggered magnetization of $\langle s_i \rangle = $ 0.0, 0.11, 0.27, and 0.32 per site, respectively. The saturation value is normalized to 0.5 per site.
}
\label{fig:fermi}
\end{figure*}

The presence of the chiral symmetry $\hat{C}$ has additional consequences for the Fermi surface, where band-crossing points are pinned. In the vicinity of a band crossing point $k^*$, one may consider a Hamiltonian $\mathcal{H}_{2\times 2} (k^*+\Delta k)$, which is projected onto the two crossing bands. The two eigenvectors must be related by $\hat{C}$, so we may choose a basis for which $\hat{C} = \sigma_x$ and $\hat{g} = e^{i k \cdot \vec{d}} \sigma_z$. Thus:
\begin{align}
    \mathcal{H}_{2\times 2} = a_y(k) \sigma_y + a_z(k) \sigma_z
\end{align}
The absence of $\sigma_x$ terms implies that the model does not support isolated, topologically protected band crossing points\cite{leonhardt2021symmetry}, leading instead to a nodal line Fermi surface depicted in Fig.~\ref{fig:fermi}(a). In the absence of magnetic order, $a_y$ is required to vanish within the glide-invariant planes, so that the nodal lines are pinned to such planes, with linked nodal chains being formed at their intersection. For any path encircling a single nodal line, the vector $(a_y,a_z)$ is required to wind an odd number of times around the origin. This chiral winding number is well-defined provided the chiral symmetry is maintained and serves as a topological invariant for the nodal lines which results in the emergence of topological modes on the surface of the material\cite{burkov2011topological,fang2015topological,bzduvsek2016nodal,schnyder2008classification}.

\begin{figure*}
\centering
\includegraphics[width=\linewidth]{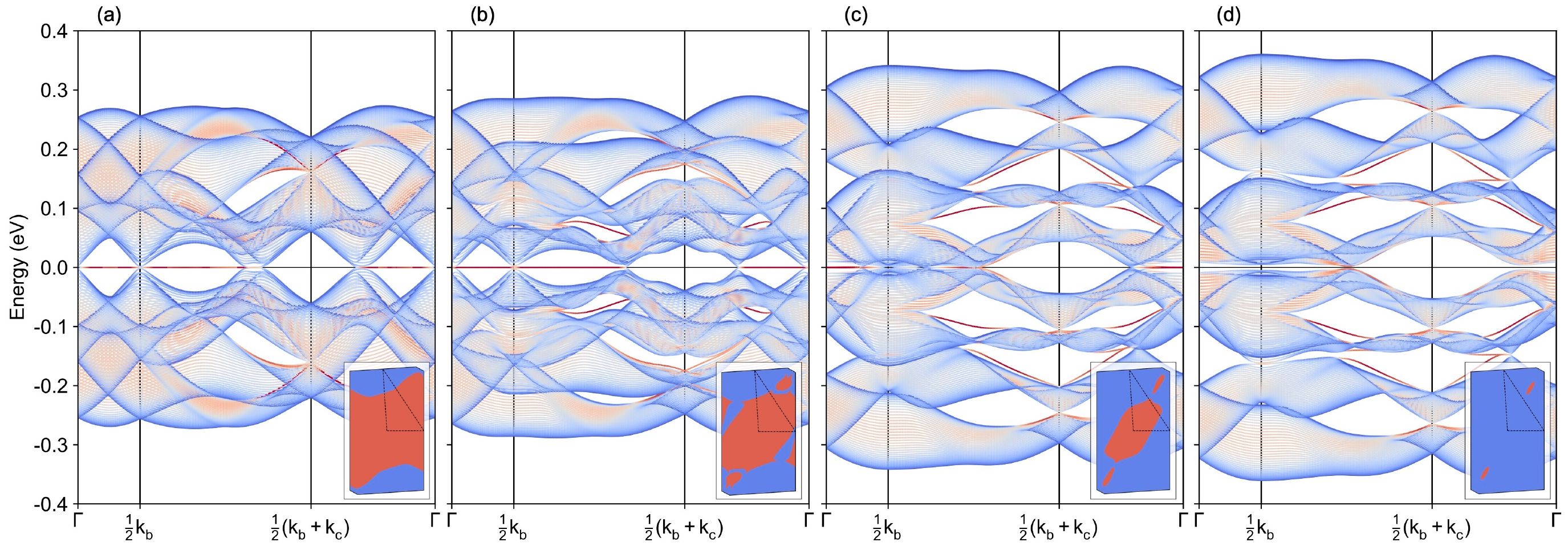}
\caption{(a)-(d): Evolution of band structure for zigzag mean field $B_{\rm ZZ}$ in the slab geometry with surfaces perpendicular to the [100] direction of the magnetic unit cell with magnitudes of $B_{ZZ}$ = 0.00, 0.04, 0.10, and 0.12 eV respectively. The locations of surface modes at the Fermi energy are depicted as insets. Bands are colored according to their projection on the surface unit cells; red bands indicate surface states. The slab calculations were performed using the \textit{topwave} python package.\cite{topwave}}
\label{fig:edges}
\end{figure*}

\subsection{Mean-Field Consequences of Magnetism}

In the following, we consider the effects of symmetry breaking magnetic order on the electronic bands in a mean-field approximation suitable to describe itinerant magnetism. In particular, we consider the effective single-particle Hamiltonian given by:
\begin{align}\label{eqn:h1p}
\mathcal{H}_{1p} = \mathcal{H}_{\rm MF} + \mathcal{H}_{\rm hop}
\end{align}
where $\mathcal{H}_{\rm hop}$ is given by Eq'n (\ref{eqn:hops}) and the finite magnetic moments are induced by the mean-field Hamiltonian:
\begin{align}
    \mathcal{H}_{\rm MF} = B_{\rm AFM} \sum_i \langle \hat{s}_i\rangle  \cdot \mathbf{c}_i^\dagger  \vec{\sigma}\mathbf{c}_i
\end{align}
where $\langle \hat{s}_i\rangle$ is a unit vector in the ordered moment direction at site $i$, and $\mathbf{c}_i^\dagger = (c_{i,+}^\dagger \ c_{i,-}^\dagger )$ creates a particle in the doublet states $|+\rangle, |-\rangle$ at site $i$ according to Eq.~(\ref{eqn:doublet1},\ref{eqn:doublet2}). Below we consider two cases: (i) ordered moment directions are set according to the predicted layered zigzag order depicted in Fig.~\ref{fig:magstructure}, and (ii) the ordered moments are oriented along the $\pm \hat{e}(\text{Y})$ axes, but instead form a nearest neighbor N\'eel order with equal magnetic and crystallographic sublattices (i.e.~up spins on Ir1 and down spins on Ir2). As discussed below, the former retains chiral symmetry, while the latter breaks it. 

It may be noted that such an approach may prove inappropriate if correlation strength is sufficiently large that IrF$_4$ is a Mott insulator in which magnetic moments are genuinely localized. However, in the absence of any experimental reports of electrical conductivity, we consider the possible mean-field effects of magnetic order in an itinerant picture. A similar approach led to prediction of a Weyl semimetal phase in pyrochore iridates\cite{wan2011topological,witczak2012topological}. It may be noted that some Ir$^{4+}$ oxides, such as Sr$_2$IrO$_4$ and Na$_2$IrO$_3$ are spin-orbit assisted Mott insulators due to the suppression of the $j_{1/2}$ bandwidth by the particular bonding geometry \cite{kim2008novel,moon2008dimensionality,witczak2014correlated}. By contrast, this effect is apparently not applicable to materials such as rutile\cite{ryden1968temperature} IrO$_2$ or some pyrochlore iridates\cite{matsuhira2011metal,tian2016field} R$_2$Ir$_2$O$_7$, which display a bonding geometry similar to IrF$_4$ and are metallic conductors. In anticipation that IrF$_4$ may be an itinerant magnet, we therefore analyze the magnetic band structure in the mean field itinerant picture. 

We first consider $\langle \hat{s}_i\rangle $ defined by the zigzag (ZZ) pattern depicted in Fig.~\ref{fig:magstructure}. For $B_{\rm ZZ} \neq 0$, the unit cell is doubled and both time reversal and glide symmetry are broken. However, because the anisotropic compass magnetism leads to a magnetic order with a layered (zigzag) antiferromagnetic pattern, a chiral symmetry is retained in the model Eq. (\ref{eqn:h1p}) for any value of $B_{\rm ZZ}$. This corresponds to $\hat{C}_{\rm mag}$: $\hat{C}\cdot\hat{T}(\vec{r}_{\rm mag})$, where $\hat{T}(\vec{r}_{\rm mag})$ is a translation between magnetic sublattices. Due to this effective chiral symmetry, the nodal lines at zero energy remain stable for finite $B_{\rm ZZ}$. However, because the magnetic field separates the bands that form nodal loops in energy, and they are no longer enforced by the non-symmorphic glide symmetry, they can be removed by contraction to zero circumference. Further, the nodal chain is broken since the nodal lines at the Fermi level are no longer pinned to high symmetry planes, and instead are free to migrate around the Brillouin zone.

In Fig.~\ref{fig:fermi}(a), we first show the $B_{\rm ZZ} = 0$ Fermi lines folded into the magnetic Brillouin zone. One can clearly see the nodal chain structure. The model also features an accidental band touching point indicated by dark yellow points. In Fig.~\ref{fig:fermi}(b), we show the effects of small symmetry breaking staggered field $B_{ZZ} = 0.04$ eV. The nodal chains are initially decoupled, forming a combination of a closed nodal ring and open nodal lines, which extend across the edges of the Brillouin zone. The accidental nodal point expands into a closed nodal ring. Since the open lines encircle the Brillouin zone, they cannot be directly contracted, and thus must merge with eachother to form closed loops. This occurs through a series of mergings at intermediate values of $B_{\rm ZZ}$. One such point is depicted in Fig.~\ref{fig:fermi}(c). Finally, for large values of the mean field $B_{\rm ZZ} \gtrsim $ 0.11 eV, the Fermi surface is composed of two isolated nodal rings (Fig.~\ref{fig:fermi}(d)), which are contracted to zero circumference and annihilate at $B_{ZZ} \approx 0.14$ eV. 

\begin{figure*}
\centering
\includegraphics[width=\linewidth]{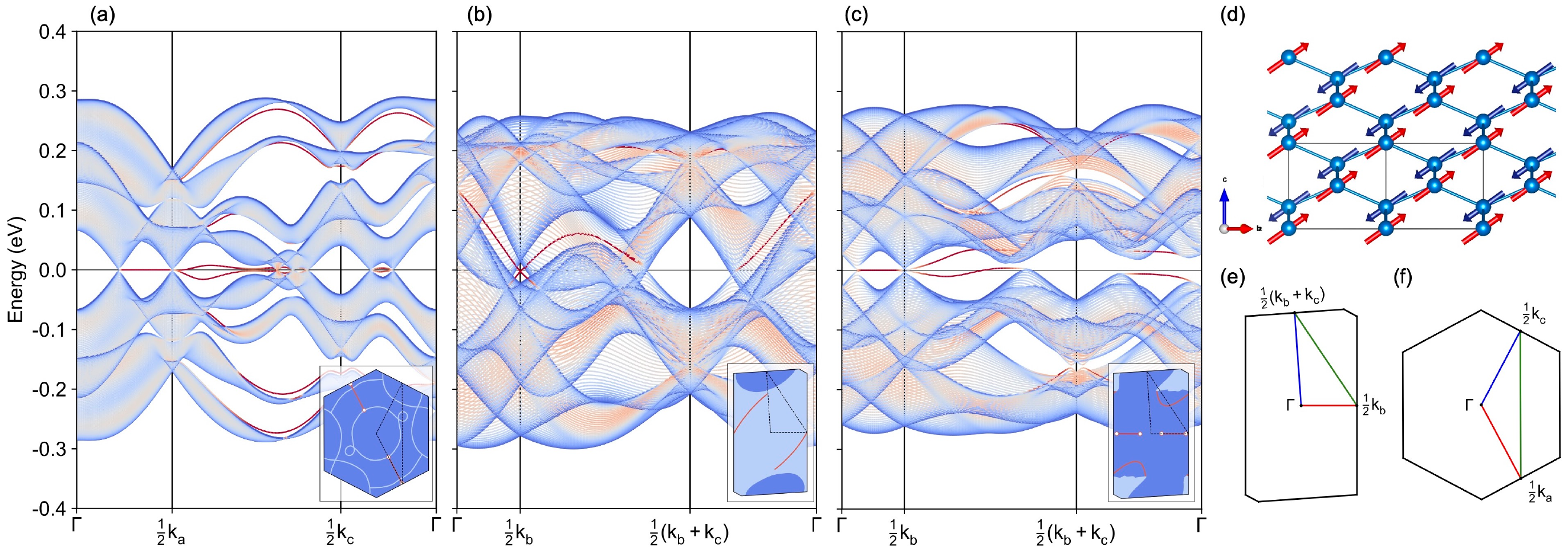}
\caption{(a) Band structure and surface modes of the [010] cut with layered zigzag magnetic ordering and $B_{ZZ}$ = 0.04 eV. (b) Band structure and surface modes of the [100] cut with next-nearest neighbor hopping value of -0.0153, and $B_{ZZ} = B_\text{N\'eel}$ = 0. (c) Band structure and surface modes of the [100] cut with Néel magnetic ordering and $B_\text{N\'eel}$ = 0.04. In the inset surface mode diagrams dark blue indicates gapped points, light blue indicates bulk gapless points, and red indicates gapless surface states within the two dimensional brillouin zone. Red circles indicate Weyl points. (d) Néel magnetic structure (e-f) Two dimensional Brillouin zones with k-points and k-paths used in the band structure diagrams labelled. The slab calculations where performed using the \textit{topwave} python package.\cite{topwave}}
\label{fig:edges2}
\end{figure*}

Given the retention of chiral symmetry with $B_{ZZ}$, it is expected that dispersionless topological drumhead surface states persist at the mean field level within the magnetically ordered phase. In order to demonstrate this effect, we show in Fig.~\ref{fig:edges}, the evolution of the band structure in the 2D Brillouin zone obtained for a slab geometry with surfaces perpendicular to the [100] direction of the magnetic unit cell. Gapless drumhead surface states appear at 2D surface $k$-points that are projections of the interiors of the bulk nodal loops onto the surface Brillouin zone. Interestingly, since the chiral operator $\hat{C}_{\rm mag}$: $\hat{C}\cdot\hat{T}(\vec{r}_{\rm mag})$ involves translation, it may be broken by surface terminations where there does not exist a $\vec{r}_{\rm mag}$ parallel to the surface. This is demonstrated for a surface perpendicular to the [010] direction in Fig.~\ref{fig:edges2}(a). In this case, the bulk spectrum is preserved and surface states persist. However, they are not strictly protected, and are no longer pinned to zero energy. The surface modes become dispersive, and intersect the Fermi energy along arcs linking different points on the surface projections of the bulk nodal lines. In this sense, $\hat{C}_{\rm mag}$ can be viewed as a \emph{quasi-symmetry} for such surface termination; as long as the field-induced splitting is small compared to the size of the gap the modes are located in, the surface states survive. 

In order to contrast these results with a chiral symmetry breaking magnetic order, we also consider the N\'eel ordering pattern depicted in Fig.~\ref{fig:edges2}(d). For this case, all relevant point group symmetries are broken, and no chiral symmetry remains. As shown in Fig.~\ref{fig:edges2}(c), at finite $B_\text{N\'eel}$, the bulk band crossing points acquire a gap, so that nodal lines do not persist even at finite energy. The Fermi surface expands into electron and hole pockets of finite volume. Elsewhere, the bulk acquires a gap except at four isolated Weyl points. The fate of the drumhead surface states is to split, and gap out essentially everywhere except Fermi arcs linking the bulk Weyl points along the path $\Gamma - \frac{1}{2}k_b$. In this way, surface states descended from the drumhead modes survive for small $B_\text{N\'eel}$ (the Weyl points merge an annihilate at $B_\text{N\'eel} \approx 0.8$ eV), but the essential structure of the paramagnetic bands is not preserved. 

Finally, we also considered the explicit breaking of the chiral symmetry in the bulk by including a second neighbor hopping consistent with the symmetry of the lattice $t_2 \sum_{\langle\langle ij \rangle\rangle}\mathbf{c}_i^\dagger \mathbf{c}_j$ (instead of an antiferromagnetic mean field), as shown in Fig.~\ref{fig:edges2}(b). For this purpose, we use $t_2 = -0.0153$ eV, which is consistent with the values in Ref.~\onlinecite{bzduvsek2016nodal}. In this case, both the bulk spectrum and edge states are strongly perturbed. The main effect is to push the bulk band crossing points away from the Fermi energy, leading to an expansion of the nodal line Fermi surface into one of finite volume. Similar to the case of chiral symmetry breaking surface termination, the surface states descended from the drumhead modes are preserved, but become dispersive. The modification of the bulk spectrum allows the majority of such descendant surface states to be pushed above the Fermi energy, except along arcs in the 2D Brillouin zone that are embedded within the gapless modes of the bulk.

\section{Conclusions}

In this work, we have shown IrF$_4$ to be a $j_{\rm eff} = 1/2$ magnet with strongly anisotropic and bond-dependent effective magnetic couplings owing to the spin-orbital composition of the local moments. The resulting magnetic interactions approximate a tetrahedral compass model on a diamond lattice, which yields an unconventional layered zigzag antiferromagnetic order. This order may be contrasted with conventional two-sublattice N\'eel order, which would typically be found for materials with bipartite lattices, dominant nearest neighbor interactions, and weak spin-orbit coupling. While we anticipate that IrF$_4$ may be a Mott insulator, we have shown that the mean-field effects of layered zigzag order preserves many of the essential topological aspects of the weakly interacting paramagnetic band structure, and retains dispersionless drumhead surface states. This is because the zigzag order preserves a chiral sublattice quasi-symmetry of the paramagnetic phase. By contrast, strong breaking of this sublattice symmetry leads to alternate phenomenology of Weyl points and Fermi arc surface states. \\

\begin{acknowledgments}
We acknowledge support through pilot funding provided by the Center for Functional Materials, Wake Forest University. Computations were performed on the Wake Forest University DEAC Cluster, a centrally managed resource with support provided in part by the University. NH acknowledges financial support from the Max Planck Institute for Solid State Research in Stuttgart, Germany.
\end{acknowledgments}

\bibliography{references}

\appendix
\section{Full Hopping}
\label{sec:appendix}
In Table \ref{tab:hoppings_rel}, we provide the scalar-relativistic hopping parameters obtained from FPLO for the bond pictured in Fig.~\ref{fig:bond}(a). For the purpose of computing the magnetic couplings, we employ hoppings from fully relativistic calculations, which effectively includes spin-orbit coupling localy, but does not strongly modify the intersite hoppings. The magnetic couplings depend most strongly on the hoppings between the occupied $t_{2g}$ orbitals. 

\begin{table}
    \caption{Scalar-Relativistic hopping parameters (in eV) for the bond pictured in Fig.~\ref{fig:bond}(a). Columns label orbitals on Ir1, and rows label orbitals on Ir2. Coordinates refer to the local cubic coordinates shown in Fig.~\ref{fig:bond}(a).
    \label{tab:hoppings_rel}}
\centering\def\arraystretch{1.1}
    \begin{ruledtabular}
    \begin{tabular}{R|RRRRR}
     & d_{yz} & d_{xz} & d_{xy}  & d_{z^2} & d_{x^2\text{-}y^2}\\
    \hline
   d_{yz}  	&	 0.151&		 -0.082&		-0.029&		 -0.131 &0.019\\
d_{xz} 	    & 	 0.028&		 -0.032&		-0.018&		-0.045&		-0.017 \\
d_{xy}  	&-0.078&		 0.014&		 0.044&		-0.300&		 0.005 \\
d_{z^2} 	& -0.029&		 -0.126&		 0.009&		 0.179&		0.012\\
d_{x^2\text{-}y^2}  	& -0.084&		 -0.257&		0.026&		0.313&		 0.034 \\

    \end{tabular}
    \end{ruledtabular}
\end{table}

\end{document}